\documentclass[11 pt, reqno]{article}

\usepackage{amsmath,amsfonts,amssymb,amsthm}
\usepackage[colorlinks=true,hyperindex=true]{hyperref}
\usepackage{cancel,bbm}
\usepackage{mathabx} 
\usepackage{comment}
\usepackage{enumitem}
\usepackage{cite}

\usepackage{chngcntr}
\counterwithin*{equation}{section}

\usepackage[a4paper, left=2cm, right=2cm, top=2 cm, bottom= 2 cm]{geometry}
\usepackage{color}
\usepackage{fancyhdr}
\usepackage{latexsym}

\newtheorem{ccounter}{ccounter}[section]
\newtheorem{thm}[ccounter]{Theorem}
\newtheorem{lem}[ccounter]{Lemma}
\newtheorem{cor}[ccounter]{Corollary}
\newtheorem{defn}[ccounter]{Definition}
\newtheorem{prop}[ccounter]{Proposition}

\def\bet{\begin{thm}}
\def\eet{\end{thm}}
\def\bel{\begin{lem}}
\def\eel{\end{lem}}
\def\bec{\begin{cor}}
\def\eec{\end{cor}}
\def\bed{\begin{Definition}}
\def\eed{\end{Definition}}
\def\bep{\begin{prop}}
\def\eep{\end{prop}}
\def\beq{\begin{equation}}
\def\eeq{\end{equation}}
\def\proof{\noindent {\bf Proof.}\ \ }
\def\bea{\begin{align*}}
\def\eea{\end{align*}}

\def\rr{\mathbb{R}}

\def\cc{\mathbb{C}}
\def\zz{\mathbb{Z}}

\def\i{\mathrm{i}}
\def\e{\rm{e}}
\def\bra{\langle}
\def\ket{\rangle}

\def\d{\mathrm{d}}
\def\e{\mathrm{e}}
\newcommand{\slim}{\mathop{\mathrm{s-lim}}\limits}
\def\Ran{{\rm Ran}\,}
\def\ac{\mathrm{ac}}

\def\Im{\mathrm{Im}}

\def\Cn0{\mathcal{C}_{n}}

\def\mn1l{m_{n-1}^{(l)}}
\def\dto{\downarrow}
\def\eps{\epsilon}

\def\H{\mathcal{H}}
\def\hatH{\widehat{\mathcal{H}}}

\def\mfe{\mathfrak{e}}

\def\Lap{\Delta}
\def\dzero{\delta_0}
\def\Hinf{H_\infty}

\def\rholac{\rho_{l, \mathrm{ac}}}
\def\rhorac{\rho_{r, \mathrm{ac}}}

\def\chixlr{\chi_x^{(l/r)}}
\def\chilrx{\chi_x^{(l/r)}}
\def\Rinf{R_\infty}
\def\Spl{R^{(s)}_l}
\def\Dyl{R^{(d)}_l}

\def\Aalp{A_\alpha}
\def\Ainf{A_\infty}
\def\A{A}
\def\wpma{w_{\pm, \alpha}}
\def\Aalpj{A_{\alpha}^{(j)}}
\def\Halpj{\mathcal{H}_\alpha^{(j)}}
\def\muaj{\mu_{\alpha, j}}
\def\sa{s_\alpha}
\def\wma{w_{-, \alpha}}
\def\wpa{w_{+, \alpha}}

\def\red{\textcolor[rgb]{0.51,0.00,0.00}}

\begin{document}

\title{Reflection probabilities of one-dimensional Schr\"odinger operators and scattering theory}
\author{Benjamin Landon$^{1}$, Annalisa Panati$^{2, 3}$,  Jane Panangaden$^{2}$, 
Justine Zwicker$^{2}$
\\
\\
$^{1}$Department of Mathematics \\
Harvard University \\
Cambridge, MA \\ \\
$^{2}$Department of Mathematics and Statistics \\
McGill University \\
Montreal, QC \\ \\
$^{3}$Aix-Marseille Universit\'e, CNRS, CPT, UMR 7332, Case 907, 13288 Marseille, France \\
Universit\'e de Toulon, CNRS, CPT, UMR 732, 83957 La Garde, France \\
FRUMAM\\ }
\maketitle
{\small
\noindent {\bf Abstract.}  The dynamic reflection probability of \cite{DaSi} and the spectral reflection probability of \cite{gesztesy1997one, gesztesy1997inverse} for a one-dimensional Schr\"odinger operator $H = - \Delta + V$ are characterized in terms of the scattering theory of the pair $(H, \Hinf)$ where $\Hinf$ is the operator obtained by decoupling the left and right half-lines $\rr_{\leq 0}$ and $\rr_{\geq 0}$.  An immediate consequence is that these reflection probabilities are in fact the same, thus providing a short and transparent proof of the main result of \cite{BRS}.  This approach is inspired by recent developments in non-equilibrium statistical mechanics of the electronic black box model and follows a strategy parallel to \cite{JLP}. }

\section{Introduction}

One-dimensional Schr\"odinger operators $H = - \Delta + V$ describe a quantum particle on $\rr$ under the influence of a potential $V$. The transmission and reflection properties of such operators concern the behaviour of wave packets originating on, say, the far left of the system and describe whether the packet is transmitted through the system to the right, or reflected back to the left.

The dynamical characterization of transmission and reflection, originating in \cite{DaSi}, uses the solution of the time dependent Schr\"odinger equation to define states asymptotically concentrated on the left or right in the distant future or past. The notions of reflection and transmission coefficients then enter through the decomposition of a state concentrated asymptotically on the left in the distant past into the superposition of two states concentrated on the left and right in the distant future.

The spectral characterization of transmission and reflection is given in terms of two special solutions of the time independent Schr\"odinger equation. The reflected and transmitted waves are obtained by expanding one solution as a superposition of the other and its conjugate (for the energies under consideration, each solution is not the multiple of a real-valued solution); in the mathematical literature, this approach originates in\cite{gesztesy1996uniqueness, gesztesy1997inverse}.

Our contribution is to characterize both of these notions of reflection and transmission in terms of the scattering theory of the pair $(H, H_\infty )$ where $H_\infty$ is the direct sum of the restrictions of $H$ to the left and right half-lines, with Dirichlet boundary conditions.  The scattering theory of this pair provides yet another notion of reflection and transmission through the elements of the scattering matrix. In fact, it will be immediate that all three notions coincide thus providing a short and transparent proof of the main result of \cite{BRS}.

The consideration of the decoupled operator $H_\infty$ appears somewhat unnatural here.  This approach originates in studies of the nonequilibrium statistical mechanics of the electronic black box model. There, the relevent transport phenomena are computed using an associated one-dimensional Jacobi matrix acting on $\ell^2 (\zz)$, where the left and right half-spaces correspond to two semi-infinite thermal reservoirs which are initially uncoupled and at thermal equilibrium. The study of the scattering theory of the decoupled and coupled operators thus enters quite naturally. We refer the interested reader to \cite{JLPi} for a complete discussion.

This approach to reflection and transmission was first developed for Jacobi matrices \cite{JLP} and later extended to the five-diagonal unitary CMV matrices in \cite{CLP}. In all of these cases, the question of whether the dynamical and spectral definitions of reflection and transmission coincide was already answered in \cite{BRS}.

The remainder of the paper is organized as follows. In the next section we introduce the different notions of reflection and transmission and state our main results. In the third and final section we provide an elementary derivation of the scattering matrix for the pair $(H, H_\infty)$.

\noindent{\bf Acknowledgements: } The authors thank V. Jak\v{s}i\'c for useful and enlightening discussions.  The work of B.L. is partly supported by NSERC.
The research of A.P. was partly supported by  ANR (grant 12-JS01-0008-01).

\section{Main results}

\subsection{Preliminaries}
We consider Schr\"odinger operators
\beq
H := - \Lap + V
\eeq
on $\rr$. We assume that $V_+ \in L^1_{\mathrm{loc}} ( \rr )$ and its negative part $V_-$ obeys $  Q(-\Lap)\subseteq Q(V_{-})$ and
\beq
\bra \varphi , V_-  \varphi \ket \leq \alpha \bra \varphi , - \Lap \varphi \ket + \beta \bra \varphi , \varphi \ket
\eeq
for some $\alpha < 1$ and all $\varphi \in Q (-\Lap )$.  We assume that $V$ is in the limit point case at $\pm \infty$.  Under these assumptions $H$ defines a self-adjoint operator bounded from below. For our purposes it will be no loss of generality to assume 
\beq
H \geq 0 .
\eeq

  We define now
\beq
H_{l/r} := -\Lap_{\infty, l/r} + V
\eeq
to be the restriction of $H$ to the left and right half spaces $L^2 (  \rr_{l/r} , \d x)$ where we denote $\rr_l = ( - \infty, 0]$ and $\rr_r = [0,\infty)$. Here, $-\Lap_{ \infty, l/r}$ is the Laplacian with Dirichlet boundary conditions on the left/right half-space.

We summarize now some of the elements of the Weyl-Titschmarsh theory (our notation is based on Appendix A of \cite{gesztesy1996uniqueness}; complete proofs may be found in the references there, or in \cite{T}).  For every $z \in \cc \backslash \rr$ there are unique solutions $u_{l/r} (z)$ of the ODE
\beq
H u = z u, \qquad u (0) =1
\eeq
that are in $L^2 ( \rr_{l/r} , \d x)$.  The Weyl $m$-functions are defined by
\beq
m_{l/r} (z) = u'_{l/r} (z, 0)
\eeq
and are Herglotz functions admitting the representation, for some real constant $a_{l/r}$,
\beq
m_{l/r} (z) = a_{l/r} + \int_{\rr} \left[ \frac{1}{ \lambda - z} - \frac{\lambda}{1 + \lambda^2} \right] \d \rho_{l/r} (\lambda)
\eeq
for a measure $\rho_{l/r}$ which satisifes
\beq
\int_\rr \frac{\d \rho_{l/r} (\lambda )}{1 + \lambda^2} < \infty.
\eeq
 In fact, it follows from \cite{gesztesy1995rank, gesztesy1996uniqueness, T} that $\rho_{l/r}$ is the spectral measure for $H_{l/r}$ and $\delta_0'$.  Here, $\delta_0' \in \H_{-2} ( H_{l/r} )$ where $\H_s ( H_{l/r})$ is the scale of spaces associated to $H_{l/r}$ and the associated spectral measure is defined as in \cite{simon1995spectral}.  Being Herglotz functions, the boundary values
\beq
\Im [m_{l/r} ( \lambda  \pm \i 0 )] :=  \lim_{\eps \dto 0} \Im [ m_{l/r} ( \lambda \pm \i \eps ) ]
\eeq
exist and are finite for Lebesgue a.e. $\lambda$. Moreover,
\beq
\Im [ m_{l/r} ( \lambda + \i 0) ] = \frac{ \d \rho_{l/r, \ac} }{ \d \lambda } ( \lambda )
\eeq
is the Radon-Nikodyn derivative of the a.c. part of $\rho_{l/r}$. 

Lastly, we define the Green's function by
\beq
G_{xy} ( z) =  \bra \delta_x , ( H - z)^{-1} \delta_y \ket .
\eeq

\subsection{Reflection probabilities}

We discuss first the spectral reflection probability of \cite{gesztesy1997one, gesztesy1997inverse}.  For any $\lambda \in \rr$ s.t. $m_{l/r} ( \lambda + \i 0)$ exists, we define the two solutions of
\beq
H \psi_{l/r} = \lambda \psi_{l/r}
\eeq
by requiring
\beq
\psi_{l/r} ( \lambda , 0) =1 , \qquad \psi_{l/r}' ( \lambda, 0 ) = m_{l/r} ( \lambda + \i 0 ). \label{eqn:bcs}
\eeq
Let $S_{l/r}$ be an essential support of the a.c. spectrum for $H_{l/r}$.  For a.e. $\lambda \in S_l$ we have that $\psi_l ( \lambda ) $ is not a multiple of a real solution and is therefore linearly independent of $\bar{ \psi}_l ( \lambda)$.  For such $\lambda$ we can expand 
\beq
\psi_r ( \lambda ) = A ( \lambda ) \bar{\psi}_l ( \lambda ) +  B ( \lambda ) \psi_l ( \lambda ).
\eeq

\begin{defn}
For a.e. $\lambda \in S_l$ the spectral reflection coefficient (relative to incidence from the left) is defined by 
\beq
\Spl ( \lambda ) := \frac{A ( \lambda ) }{ B ( \lambda )}.
\eeq
and the spectral reflection probability is defined by $ |\Spl ( \lambda ) |^2$.  For $\lambda \notin S_l$ set $ \Spl ( \lambda) =1$.  If $\Spl ( \lambda ) = 0$ a.e. on a Borel set $\mfe$ then $H$ is said to be spectrally reflectionless on $\mfe$.
\end{defn}
Using (\ref{eqn:bcs}) one obtains for $\lambda \in S_l$,
\beq
\Spl ( \lambda) = \frac{ m_r ( \lambda + \i 0 ) - m_l ( \lambda - \i 0 ) }{ m_r ( \lambda + \i 0  ) - m_l ( \lambda + \i 0 ) } .
\eeq

We now discuss the dynamical reflection probability of \cite{DaSi, BRS}.  Let $j_x \leq 1$ be a smooth function that is $0$ on $(-\infty , x]$ and $1$ for $[x+1, \infty)$.  The strong limits
\beq
P^\pm_{r} := \slim_{t \to \pm \infty} \e^{ \i t H} j_x \e^{-\i t H } P_{\ac} ( H), \qquad P^\pm_{l} := \slim_{t \to \pm \infty} \e^{ \i t H} (1 -j_x) \e^{-\i t H } P_{\ac} ( H), 
\label{eqn:Ppm existence}
\eeq
exist and do not depend on the choice of $x$. The operator $P_{l/r}^\pm$ is the projection onto the space $\H_{l/r}^\pm$ where
\beq
\H_l^{\pm} := \left\{ \varphi \in \H_{\ac} ( H) : \forall x, \lim_{t \to \pm \infty} || j_x \e^{ - \i t H} \varphi || = 0 \right\}
\eeq
is the set of states concentrated asymptotically on the left in the distant future/past, and $\H_{r}^\pm$ is given by a similar definition with $j_x \to 1 - j_x$.  Davies and Simon proved the following:

\bet[Theorem 3.3 of \cite{DaSi}] \label{thm:dasa} Under our hypotheses on $V$ we have the following decomposition for the absolutely continuous subspace of $H$.
\beq
\H_{\ac} ( H ) = \H_l^+ \oplus \H_r^+ = \H_l^- \oplus \H_r^- .
\eeq
\eet

Note by time reversal invariance, $\H^\pm_{l/r} = \widebar{ \H_{r/l}^\mp}$. We will see later that the restriction of $H$ to $\H_{l/r}^\pm$ has simple spectrum. As each of the spectral projections commutes with $H$, the operator $P_l^+ P_l^- P_l^+$ acts as an operator of multiplication by a function we denote by $| \Dyl  ( \lambda )|^2$. This function is defined for a.e. $\lambda$ in an essential support of the a.c. spectrum of the restriction of $H$ to $\H_l^+$.  We will see later that this is precisely $S_l$.  Following \cite{BRS} we have the following definition.
\begin{defn} \label{def:dynref}
The dynamical reflection probability is defined for a.e. $\lambda \in S_l$ by $| \Dyl ( \lambda)|^2$ and by $1$ for $\lambda \notin S_l$.  If $|\Dyl|^2$ vanishes a.e. on a Borel set $\mfe$ then we say that $H$ is dynamically reflectionless on $\mfe$.
\end{defn}
\noindent{\bf Remark.} In \cite{BRS} $H$ is said to be dynamically reflectionless on a Borel $\mfe \subseteq \rr$ if $P_\mfe (H) P_l^+ = P_\mfe (H) P_r^-$  (where $P_\mfe (H)$ is the spectral projection onto $\mfe$) and $\mfe$ is contained up to a set of measure $0$ in an essential support of the a.c. spectrum of $H$. We will see later that $| \Dyl ( \lambda ) |^2 \neq 1$ only on $S_l \cap S_r$ and on this set the operator $P_r^+ P_r^- P_r^+$ acts by multiplication by the same function $ | \Dyl ( \lambda ) |^2$.  It follows that this definition coincides with ours (these observations have reduced the question of the equivalence of our definition and that of \cite{BRS} to the fact that for two orthogonal projections we have $P+Q=1$ iff $PQP = 0$ and $(1- P)(1 - Q) (1- P) = 0$).

As we will see, the link between the dynamical and spectral notions of reflection is provided by scattering theory.  To be more precise, we will prove:
\bet \label{thm:main}
The spectral and dynamic reflection probabilities are identical,
\beq
| \Spl ( \lambda ) |^2 = | \Dyl ( \lambda ) |^2.
\eeq
Moreover, they are given by any diagonal element of the scattering matrix of the pair $(H, \Hinf )$, which will be defined below. Consequently, $H$ is dynamically reflectionless on a Borel set $\mfe$  iff it is spectrally reflectionless on $\mfe$.
\eet

\subsection{Scattering theory of decoupled Schr\"odinger operators}
\label{sbsec:sacttering}
In this section we review the elements of scattering theory we need and provide a proof of Theorem \ref{thm:main}. Recall our definition of $H_\infty$ as the direct sum of the restriction of $H$ to the right and left half-lines.  For the difference of the resolvents we have the formula \cite{simon1995spectral}
\beq \label{eqn:resform}
( H - z )^{-1} - ( \Hinf - z)^{-1} = G_{00} (z)^{-1} \vert (H - z)^{-1} \dzero \rangle \langle (H - z)^{-1} \dzero \vert 
\eeq
which holds in particular for $z = -1$.  Recall that $\delta_0 \in \H_{-1} (H)$ and so $(H-z)^{-1} \delta_0$ is a bonafide element of $\H$.  From \eqref{eqn:resform} we see immediately that the wave operators \cite{RS3}
\beq
w_{\pm} := \slim_{t \to \pm \infty} \e^{ \i t H} \e^{ - \i t \Hinf } P_{\ac} ( \Hinf )
\eeq
exist and are complete (complete meaning that $\Ran w_{\pm} = \H_{\ac} ( H )$).

The scattering matrix
\beq
s := w_+^* w_-
\eeq
is a unitary operator taking $\H_{\ac} ( \Hinf )$ to itself. Moreover, it commutes with $\Hinf$ and hence acts as multiplication by a $2 \times 2$ matrix which we denote by $s ( \lambda )$ on
\beq
\H_{\ac} ( \Hinf ) = \H_{\ac} ( H_l ) \oplus \H_{\ac} ( H_r ) \cong L^2 ( \rr , \d \rholac ) \oplus L^2 ( \rr , \d \rhorac ). \label{eqn:hac}
\eeq
Let us elucidate the above identification.  First, the spectral measure for $\delta_0'$ and $H_{l/r}$ is defined as in \cite{simon1995spectral}.  Secondly, $\delta_0'$ is cyclic for $H_{l/r}$ in the sense that the linear span of 
\beq
\left\{  ( H_{l/r}-z)^{-1} \delta_0' : z \in \cc \backslash \rr \right\}
\eeq
is dense in $L^2 ( \rr_{\leq 0}, \d x) / L^2 ( \rr_{\geq 0}, \d x)$ \cite{gesztesy1995rank, gesztesy1996uniqueness, T}.  From this the identification in \eqref{eqn:hac} follows.

We now give a formula for the scattering matrix which is immediate from standard stationary scattering theory \cite{Y}; we include an elementary derivation in Section \ref{sec:scatder} for completeness.  Before we state the formula we apply a unitary transformation taking the space on the RHS of (\ref{eqn:hac}) to
\beq
 \hatH := L^2 ( \rr, \eta_{l} \d \lambda ) \oplus L^2 ( \rr, \eta_{r} \d \lambda )
\eeq
where $\eta_{l/r}$ is the characteristic function of $S_{l/r}$ (the transformation is given by multiplication by the square root of the appropriate Radon-Nikodyn derivative).  The formula we give is for the scattering matrix acting on this transformed space.  This transformation is essentially notational and is standard in scattering theory; with this convention, the scattering matrix is a unitary $2\times 2$ matrix for every $\lambda$. It is clear that the value of the diagonal elements of the scattering matrix are unaffected.

\bet  \label{thm:scform} The scattering matrix for the pair $( H, \Hinf )$ acting on $\hatH$  is given by
\beq
s_{ab} ( \lambda ) = \delta_{ab} + 2 \i G_{00}  ( \lambda + \i 0 ) \sqrt{ \Im [ m_a ( \lambda + \i 0 ) ]  \Im [ m_b ( \lambda + \i 0 ) ]  }
\eeq
for $a, b \in \{ l, r \}$. 
\eet
Note that by unitarity of the scattering matrix we have $| s_{ll} ( \lambda ) |^2 = | s_{rr} ( \lambda ) |^2$ and by the above formula both are  identically $1$ except for $\lambda \in S_l \cap S_r$.

From the point of view of scattering theory it is natural to introduce $| s_{ll} ( \lambda ) |^2$ as the reflection probability and introduce the definition of reflectionless operators as in the 
Jacobi and CMV cases. The following proof shows that all of these definitions coincide.

\noindent{\bf Proof of Theorem \ref{thm:main}.} From the formula \cite{gesztesy1996uniqueness}
\beq
G_{00} ( z ) = \frac{-1}{ m_l ( z ) + m_r ( z) }
\eeq
we see
\beq
| \Spl ( \lambda ) |^2 = | s_{ll} ( \lambda )|^2 .
\eeq
Next we claim that we can replace the smooth cut-off function $j_a$ in the definition of the asymptotic projections.  Let $\chi_0$ be the characteristic function of $[0, \infty)$.  We claim that
\beq
\slim_{t \to \infty} \e^{\i t H} (j_x - \chi_0) \e^{ - \i t H} P_{\ac} (H) = 0.
\eeq
It suffices to prove that $||  (j_x - \chi_0) \e^{ - \i t H} P_{\ac} (H) \varphi || \to 0$ as $t \to \infty$ for $\varphi \in D(H)$.  Let $\psi = (H+1) \varphi$ and $C = (j_x - \chi_0) (H+1)^{-1}$.  Then,
\beq
||  (j_x - \chi_0) \e^{ - \i t H} P_{\ac} (H) \varphi || = || C \e^{ - \i t H} P_{\ac} \psi ||.
\eeq
By Theorem 3.2 of \cite{DaSi}, $C$ is compact (trace-class even), hence by a well-known fact the RHS then goes to $0$ as $t \to \infty$ (see Lemma 2 on page 24 of \cite{RS3}).  

We may therefore compute $P_{l/r}^\pm$ with $\chi_0^{(l/r)}$ instead of $j_a$ and $1-j_a$, where $\chi_0^{(l/r)}$ is the characteristic function of $(-\infty, 0]$ and $[0, \infty)$ respectively.  Since $H_\infty$ commutes with $\chi_0^{(l/r)}$ we immediately obtain that
\beq
P_{l/r}^\pm = w_\pm \chi_0^{l/r} w_\pm^*. \label{eqn:projform}
\eeq
As a consequence, $w_\pm$ is a unitary map from $\H_{\ac} ( H_{l/r} )$ to $\H_{l/r}^\pm$ intertwining $\Hinf$ and $H$.  From this we see that the restriction of $H$ to $\H_{l/r}^\pm$ has simple spectrum and that $S_{l/r}$ is an essential support of the a.c. spectrum for this restriction.

 The formula \eqref{eqn:projform} and the definition of the dynamical reflection probability yields
\beq
| \Dyl  ( \lambda ) |^2 = |s_{ll} ( \lambda ) |^2
\eeq
for a.e. $\lambda \in S_l$.  Since the RHS is identically $1$ outside of $S_l$ we see that the above equality holds a.e. 
A similar calculation yields that $P_r^+ P_r^- P_r^+$ acts by multiplication by the same function (recall $|s_{ll } | = | s_{rr} |$). Hence we see that $| \Dyl  ( \lambda )|^2 \neq 1$  only on $S_l \cap S_r$ justifying the remark following Definition \ref{def:dynref}. We conclude Theorem \ref{thm:main}. \qed

We would like to give a final comment on the decomposition of a $\psi \in \H_l^-$ as $\psi = P_l^+ \psi + P_r^+ \psi$.  From the above proof we see that $w_\pm \delta_0 '$ is cyclic for the restriction of $H$ to $\H_{l/r}^\pm$ (where we are implicitly thinking of $\delta_0'$ as an element of $\H_{-2} ( H_{l/r} )$).  After applying a unitary map (again, multiplication by the square root of the appropriate Radon-Nikodyn derivative) to the associated spectral representation, 
we get a representation of (the restriction of) $H$ as multiplication by $\lambda$ on $L^2 ( \rr, \eta_{l/r} (\lambda) \d \lambda )$. In this representation, $P_{l/r}^+ \varphi ( \lambda) = s_{l, l/r } ( \lambda) \varphi ( \lambda )$.  This motivates our choice of notation for the dynamical reflection probability as $| \Dyl ( \lambda ) |^2$, as we could define $\Dyl ( \lambda )$ as the coefficient of $P_l^+ \varphi$ which is now seen to be $s_{ll} ( \lambda )$.

\section{Derivation of scattering matrix} \label{sec:scatder}
In this section we give an elementary derivation of Theorem \ref{thm:scform}.  By the invariance principle \cite{RS3} the wave operators and scattering matrix of the pair $( H, \Hinf )$ is the same as that for the pair $ ( - ( H + 1)^{-1}, - ( \Hinf + 1)^{-1} )$ so we compute the latter.  For notational convenience we denote
\beq
R = - ( H + 1)^{-1}, \qquad \Rinf = - ( \Hinf + 1)^{-1}.
\eeq

We compute only $s_{ll} ( \lambda )$.  The other elements are the same.  For $f_l$ and $g_l \in \H_{\ac} ( H_l )$ we have
\begin{align}
\notag \bra f_l, ( s -1 ) g_l \ket &= \bra f_l , ( w_+^*  - w_-^* ) w_-  g_l \ket  \\
&= \lim_{t \to \infty} \bra f_l , ( \e^{ \i t \Rinf } \e^{ - \i t R } - \e^{ - \i t \Rinf } \e^{ \i t R} ) w_- g_l \ket \notag \\
&= \lim_{\eps \dto 0}  \int_\rr \i \e^{ - \eps |t| } \bra f_l ,  \e^{ \i t \Rinf }  ( \Rinf - R ) e^{ - \i t R } w_- g_l \ket \d t \label{eqn:der1}
\end{align}
where in the last line we use the fundamental theorem of calculus and then replace the integral $\lim_{t \to \infty } \int_{-t}^t$ by an Abel sum, which is allowed as the integrand is bounded and we know the limit exists.  By (\ref{eqn:resform}) and the fact that
\beq 
G_{00} ( z )^{-1} ( H - z)^{-1} \dzero = (H_l - z )^{-1} \dzero ' \oplus ( H - z)^{-1} \dzero'  \label{eqn:deltas}
\eeq
(this can be derived from the formulas in Appendix A of \cite{gesztesy1996uniqueness} in the same way that the half-line case is done in \cite{gesztesy1995rank}) we see that the integrand in the last line of \eqref{eqn:der1} is given by
\beq \label{eqn:der3}
 \i \e^{ - \eps |t| } \bra f_l ,  \e^{ \i t \Rinf }  ( \Rinf - R ) e^{ - \i t R } w_- g_l \ket = - \i     \e   ^{ - \eps |t| }    \bra f_l , \e^{ \i t \Rinf } ( H_l - z)^{-1} \dzero ' \ket \bra R \dzero , w_- \e^{ -\i t \Rinf } g_l \ket .
\eeq
We evaluate the latter inner product following an argument similar to \cite{JKP}. Let $h_l = \e^{ - \i t \Rinf } g_l$.  We have,
\begin{align}
\bra R \dzero, w_- h_l \ket &= \lim_{ t \to \infty } \bra R \dzero , \e^{ - \i t R } e^{ \i t \Rinf} h_l \ket \notag \\
&= \bra R \dzero, h_l \ket - \lim_{ \eps_1 \dto 0 } \int_0^\infty \e^{ - \eps_1 t } \i \bra R \dzero   ,  \e^{ - \i t R} (   R - \Rinf ) \e^{ \i t \Rinf } h_l \ket \d t \label{eqn:der2}
\end{align}
where we have used the FTC and replaced the integral $\lim_{t \to \infty} \int_0^t$ by its Abel sum. Using (\ref{eqn:resform}) and (\ref{eqn:deltas}) we see that the integral in the last line of \eqref{eqn:der2} equals
\beq
\int_0^\infty \e^{ - \eps_1 t } \i \bra R \dzero   ,  \e^{ - \i t R} (   R - \Rinf ) \e^{ \i t \Rinf } h_l \ket \d t =\int_{0}^\infty \i \e^{ - \eps_1 t } \bra \dzero , R \e^{ - \i t R} R \dzero \ket \left[ \int_\rr \e^{ - \i t ( \lambda + 1)^{-1} } h_l ( \lambda ) \frac{ \d \rholac (\lambda ) }{ \lambda  + 1 } \right] \d t .
\eeq
Everything here is absolutely integrable so we use Fubini and evaluate the $t$ integral yielding
\begin{align}
&\int_\rr h_l ( \lambda ) \i \int_0^\infty \bra \dzero , R \e^{ - \i t ( R + ( \lambda + 1)^{-1} - \i \eps_1 )}R \dzero \ket \d t  \frac{ \d \rhorac ( \lambda ) }{ \lambda + 1} \notag \\
= &\int_\rr h_l \bra \dzero , R ( R + ( \lambda + 1)^{-1}  - \i \eps_1 )^{-1}  R \dzero \ket \frac{ \d \rholac ( \lambda )}{ \lambda + 1 } .
\end{align}
We now concern ourselves with the limit $\eps_1 \dto 0$.  First note that
\begin{align}
\lim_{\eps_1 \dto 0 } \bra \dzero , R ( R + ( \lambda + 1)^{-1} - \i \eps )^{-1} R \dzero \ket = -G_{00} ( -1 )  + \lim_{ \eps_1 \dto 0 } \bra \dzero , (  H - \lambda- \i \eps ( H+  1 ) ( \lambda + 1) )^{-1} \dzero \ket
\end{align}
whenever the right hand limit exists.  In a moment we will prove that
\beq \label{eqn:twolimits}
 \lim_{ \eps_1 \dto 0 } \bra \dzero , (  H - \lambda- \i \eps_1 ( H+  1 ) ( \lambda + 1) )^{-1} \dzero \ket = \lim_{\eps_1 \dto 0 } \bra \dzero , ( H - \lambda - \i \eps_1 )^{-1} \dzero \ket = G_{00} ( \lambda + \i 0)
\eeq
whenever the limit on the RHS exists and is finite.  Suppose for the remainder of the proof that $g_l$ is nonzero only on a set on which  the RHS of \eqref{eqn:twolimits} converges to $G_{00} ( \lambda + \i 0)$ uniformly.  By the DCT we see that we can pass the limit $\eps_1 \dto 0$ through the integral and we obtain
\beq
\bra R \dzero , w_- \e^{ - \i t \Rinf } g_l \ket =  - \int_\rr \e^{ \i t ( \lambda + 1)^{-1} } G_{00} ( \lambda + \i 0 ) g_l ( \lambda ) \frac{ \d \rholac ( \lambda ) }{ \lambda + 1}.
\eeq
Plugging this into \eqref{eqn:der3} and using Fubini we see that
\beq
\bra f_l , ( s -1 ) g_l \ket = \lim_{\eps \dto 0} \int_\rr \int_{\rr} \bar{f}_l ( \lambda' )  g_l ( \lambda) \left[ \int_\rr \e^{ \i t  ( \lambda + 1)^{-1} - ( \lambda' + 1)^{-1}  - t | \eps | } \d t \right] \frac{ \d \rholac ( \lambda ) }{ \lambda + 1 }  \frac{ \d \rholac ( \lambda' ) }{ \lambda' + 1 } 
\eeq
To complete the computation we use the fact that
\beq
\lim_{\eps \dto 0} \int_{\rr} \e^{  \i t ( ( \lambda +1)^{-1} - ( \lambda' + 1)^{-1} ) - \eps | t| } \d t \to \delta ( (\lambda +1)^{-1} - ( \lambda' + 1)^{-1} ) = ( \lambda + 1)^2 \delta ( \lambda - \lambda' ) .
\eeq
To be more precise, since $g ( \lambda ) G_{00} ( \lambda + \i 0 ) \frac{ \d \rholac } { \d \lambda } ( \lambda ) ( \lambda + 1)^{-1}$ is in $L^1 ( \rr , \d x)$ it is not too hard to see that
\beq
\int g_l ( \lambda ) G_{00} ( \lambda + \i 0 )  \int_{\rr} \e^{  \i t ( ( \lambda +1)^{-1} - ( \lambda' + 1)^{-1} ) - \eps | t| } \d t \frac{ \d \rholac }{ \lambda + 1 }  \to g_l ( \lambda ' ) G_{00} ( \lambda' + \i 0 ) \frac{ \d \rholac }{ \d \lambda } ( \lambda' ) ( \lambda' + 1)
\eeq
strongly in $L^1$ provided that we also take $g_l$ to have compact support.  If we then chose $f_l$ so that $f_l ( \lambda' ) \frac{ \d \rholac }{ \d \lambda} ( \lambda' )$ is bounded with $f_l$ also of compact support then we can pass the above limit through the $\lambda'$ integration and obtain 
\beq
\bra f_l, ( s - 1 ) g_l \ket = -\i \int_\rr \left( \frac{ \d \rholac } { \d \lambda} ( \lambda ) \right)^2 \bar{f}_l (\lambda) g_l ( \lambda ) G_{00} ( \lambda + \i 0 ) \d \lambda .
\eeq
 We have therefore obtained the formula for the special class of $f_l$ and $g_l$ described in the proof.  The extension to all of $L^2$ is a density argument which we omit.

It remains to obtain (\ref{eqn:twolimits}).  With $\mu_0$ the spectral measure for $(H, \dzero)$ we have
\begin{align}
&|\bra \dzero , ( H - \lambda - \i \eps ( H + 1)( \lambda + 1))^{-1} \dzero \ket - \bra \dzero , ( H - \lambda - \i \eps (\lambda + 1)^2  )^{-1} \dzero \ket | \notag \\
\leq &  \int  \frac{  ( \lambda - \lambda') ( \lambda + 1)  \eps}{ | ( \lambda' - \lambda - \i \eps ( \lambda +1)^2 ) ( \lambda' - \lambda - \i \eps ( \lambda' +1)( \lambda + 1) )| } \d \mu_0 ( \lambda') \notag \\
\leq  & \int \frac{ \eps ( \lambda + 1)}{ |  \lambda' - \lambda - \i \eps ( \lambda +1)^2  |} \d \mu_0 ( \lambda') \notag \\
\leq & \int_{ | \lambda  - \lambda' | \leq \delta } \frac{ ( \lambda + 1) \eps }{ | \lambda' - \lambda - \i \eps ( \lambda +1)^2 | } \d \mu_0 ( \lambda ') + \eps C_\delta
\end{align}
for any $\delta > 0$. The last integral is bounded by
\beq
C_\lambda \sqrt{ \eps^2 + \delta^2} \Im [ G_{00} ( \lambda + \i \eps ( \lambda +1 )^2 )]
\eeq
and the claim follows.


\begin{thebibliography}{9999}
\bibitem[BRS]{BRS} Breuer, J., E. Ryckman, and B. Simon. ``Equality of the spectral and dynamical definitions of reflection." \emph{Commun. Math. Phys.} 295.2 (2010): 531-550.
\bibitem[CLP]{CLP} Chu, S., B. Landon, and J. Panangaden. ``Reflectionless CMV matrices and scattering theory. \emph{Lett. in Math. Phys.} 
 105. 4  (2015) :463-481
\bibitem[DS]{DaSi} Davies, E.B., and B. Simon. ``Scattering theory for systems with different spatial asymptotics on the left and right." \emph{Commun. Math. Phys.} 63.3 (1978): 277-301.
\bibitem[GNP]{gesztesy1997one} Gesztesy, F., R. Nowell, and W. P\"otz. ``One-dimensional scattering theory for quantum systems with nontrivial spatial asymptotics." \emph{Diff. Integral Eqs.} 10.3 (1997): 521-546.
\bibitem[GS1]{gesztesy1995rank} Gesztesy, F., and B. Simon. ``Rank one perturbations at infinite coupling." \emph{J. Funct. Analysis} 128.1 (1995): 245-252.
\bibitem[GS3]{gesztesy1996uniqueness} Gesztesy, F., and B. Simon. ``Uniqueness theorems in inverse spectral theory for one-dimensional Schrödinger operators." \emph{Trans. Amer. Math. Soc.} 348.1 (1996): 349-373.
\bibitem[GS4]{gesztesy1997inverse} Gesztesy, F., and B. Simon. ``Inverse spectral analysis with partial information on the potential, I. the case of an AC Component." \emph{Helv. Phys. Acta} 70 (1997): 66-71.
\bibitem[JKP]{JKP} Jak\v si\'c, V., Kritchevski, E., Pillet, C.-A.: Mathematical theory of the
Wigner-Weisskopf atom. In {\em Large Coulomb Systems.} J.~Derezi\'nski and H.~Siedentop editors.
Lecture Notes in Physics {\bf 695}, Springer, Berlin, 2006.
\bibitem[JLPa]{JLP} Jak\v{s}i\'c, V., B. Landon, and A Panati. ``A note on reflectionless Jacobi matrices." \emph{Commun. Math. Phys.} 332 (2014): 827-838.
\bibitem[JLPi]{JLPi} Jak\v{s}i\'c, Vojkan, Benjamin Landon, and Claude-Alain Pillet. ``Entropic fluctuations in XY chains and reflectionless Jacobi matrices." \emph{Ann. Henri Poincar\'e}. 14.7 (2013): 1775-1800.
\bibitem[RS]{RS3}Simon, B., and M. Reed. \emph{Scattering Theory}. Vol. 3. San Diego: Academic Press, 1979. Print. Modern Methods of Mathematical Physics.
\bibitem[S]{simon1995spectral} Simon, B. ``Spectral analysis of rank one perturbations and applications." CRM Lecture Notes. Vol. 8. 1995.
\bibitem[T]{T} Teschl, G. \emph{Mathematical methods in quantum mechanics}. Vol. 157. American Mathematical Soc., 2014.
\bibitem[Y]{Y} Yafaev, D.R. \emph{Mathematical scattering theory: General theory}. No. 105. American Mathematical Soc., 1998.



\end{thebibliography}
\end{document}